\documentclass[aps,prc,twocolumn,showpacs,superscriptaddress]{revtex4}
\usepackage[latin1]{inputenc}
\usepackage{bm}
\usepackage{epsfig}
\usepackage{amsthm}
\usepackage{amsfonts}
\usepackage{float}
\usepackage{amsmath,amssymb}
\usepackage{color}
\usepackage{graphicx}% Include figure files
\usepackage{dcolumn}% Align table columns on decimal point
\usepackage{bm}% bold math
\usepackage{hyperref}

\newcommand{\nn}{\nonumber}
\newcommand{\be}{\begin{equation}}
\newcommand{\ee}{\end{equation}}
\newcommand{\bea}{\begin{eqnarray}}
\newcommand{\eea}{\end{eqnarray}}

\newcommand{\om}{\omega}

\newcommand{\ov}{\overline}

\newcommand{\vk}{\bold  k}
\newcommand{\vp}{\bold p}
\newcommand{\vq}{\bold q} 
\newcommand{\vl}{\bold l}

\newcommand{\mn}{\mu\nu}
\newcommand{\del}{\partial}

\begin{document}

\title{Contribution of kaon component in viscosity and conductivity of hadronic medium} 
\author{Mahfuzur Rahaman}
%\email{}
\affiliation{Variable Energy Cyclotron Centre
1/AF Bidhannagar, Kolkata 700 064,
India}
\affiliation{Homi Bhabha National Institute, Training School Complex, Anushaktinagar, Mumbai - 400085, India}
\author{Snigdha Ghosh}
%\email{}
\affiliation{Variable Energy Cyclotron Centre
1/AF Bidhannagar, Kolkata 700 064,
India}
\affiliation{Homi Bhabha National Institute, Training School Complex, Anushaktinagar, Mumbai - 400085, India}
\author{Sabyasachi Ghosh}
\email{sabyaphy@gmail.com}
\affiliation{Department of Physics, University of Calcutta, 92, 
A. P. C. Road, Kolkata - 700009, India}
\author{Sourav Sarkar}
%\email{}
\affiliation{Variable Energy Cyclotron Centre
1/AF Bidhannagar, Kolkata 700 064,
India}
\affiliation{Homi Bhabha National Institute, Training School Complex, Anushaktinagar, Mumbai - 400085, India}
\author{Jan-e Alam}
%\email{}
\affiliation{Variable Energy Cyclotron Centre
1/AF Bidhannagar, Kolkata 700 064,
India}
\affiliation{Homi Bhabha National Institute, Training School Complex, Anushaktinagar, Mumbai - 400085, India}
%

%\maketitle

\begin{abstract}
%The two-point correlation functions of kaonic viscous stress tensors and electro-magnetic 
%currents have been calculated respectively to estimate the contributions of the strange sector in the  
%shear viscosity and electrical conductivity of hadronic medium. 
%In the one-loop correlators, kaon propagators contain
%a non-zero thermal width that leads to non-divergent values of transport coefficients.
%
With the help of effective Lagrangian densities of strange hadrons, we have calculated
kaon relaxation time from several loop and scattering diagrams at tree-level,
which basically represent contributions from $1\leftrightarrow 2$ and $2\leftrightarrow 2$ type of collisions.
Using the total relaxation time of kaon, shear viscosity and electrical conductivity
of this kaon component have been estimated. The high temperature, close to transition
temperature, where kaon relaxation time is lower than life time of RHIC or LHC matter,
may be the only relevant domain for this component to contribute in hadronic dissipation. 
Our results suggest that kaon can play an important role in the enhancement of shear viscosity
and electrical conductivity of hadronic matter near the transition temperature.
%
%With the help of effective Lagrangian densities of strange hadrons, we have calculated
%in-medium self-energy of kaon for different possible mesonic loops, whose imaginary
%part provide the estimation of kaon thermal width. 
%
%It is observed that near the quark-hadron 
%transition temperature, the contribution of kaons
%to shear viscosity is larger and increases faster with temperature than pionic contribution.
%In case of electrical conductivity the trend due kaon component  
%appears to be opposite in nature with respect to pion component. 
\end{abstract}
%
%\pacs{11.10.Wx,12.39.Ki,21.65.-f,51.20+d,51.30+i}
%
\maketitle
%
%
% 
%%%%%%%%%%%%%%%%%%%%%%%%%%%%%%%%%%%%%%%%%%%%%%%%%%%%%%%%%%%%%%%%%%%%%%%%%%%
\section{Introduction}
\label{sec:intro} Collisions of heavy ions at highly relativistic energies at 
the Relativistic Heavy Ion Collider (RHIC)~\cite{Arsene,Adcox,Back,Adams} 
and Large Hadron Collider (LHC)~\cite{Aamodt1,Aamodt2,Caffarri} 
produce a very hot and dense matter with quarks and gluons as its elementary 
constituents; this state of matter is known as quark-gluon plasma (QGP). 
It behaves like a perfect fluid characterized by low shear viscosity to entropy ratio \cite{Arsene,Adcox,Back,Adams}. 
Within the framework of  ADS/CFT correspondence \cite{Policastro} Kovtun, Son and Starinets (KSS) 
\cite{Kovtun}, conjectured the lowest bound of shear viscosity ($\eta$) as,  $\eta\geq s/4\pi$, 
where $s$ is the entropy density.  One of the major objectives of 
these experiments  is to understand the quark-hadron transition in the early universe. 

In this context it is important to understand how the matter, created in the experiments of 
heavy ion collision, evolves in space and time. The relativistic viscous hydrodynamics 
is an efficient tool to simulate this evolution.   
Various transport coefficients such as shear and bulk viscosities
are required as inputs to these simulations in addition to initial conditions and equation of state (EoS).  
These transport coefficients of quark and hadronic matter can be calculated microscopically by using
effective interaction  models and this is one of the active fields of contemporary research 
in heavy ion physics. 
It is important to investigate the dependence of these transport coefficients on temperature of the 
medium to characterize the state of the matter.
%
%For example,  it has been  demonstrated that at the critical point 
%$\eta /s$ and $\zeta/s$ achieve minimum and maximum respectively ($\zeta$ is the bulk viscosity). 
%The non-zero viscosity is also required to explain data related to elliptical flow ($v_2$) of hadrons~\cite{Schenke}, 
%diffusion of heavy quarks, suppression of jets and transverse momentum 
%correlations  of the dissipative  matter created after the collision. 
In particular, it is crucial to know the temperature dependence of $\eta/s$ in order to describe the elliptic flow of hadrons in ultra-relativistic heavy-ion collisions at RHIC and LHC \cite{Niemi,Rischke1}.
%

%
%Two basic formalisms are used to investigate microscopically the transport coefficients of 
%non-equilibrium systems: 
%(1) the relativistic Boltzmann transport equation (BTE) 
%and (2) the Kubo formalism. The BTE has been solved with in relaxation time approximation with 
%thermal cross-sections of colliding particles  as inputs,
%whereas spectral function of the basic degrees of freedom 
%is used in Kubo formalism.  The optical theorem connects the two approaches. In the present paper, 
%we have used the Kubo formalism in order to find out the shear viscosity of hadron gas.  
%

A lot of work have been done on calculation of the shear viscosity of 
%quark matter\cite{Arnold,Nills,Meyer_eta}, % 
hadronic matter \cite{Itakura,Dobado,Nicola,Weise1,SSS,Ghosh_piN,Gorenstein,HM,Hostler,Khvorostukhin}. 
Some effective QCD model calculations~\cite{Purnendu,Redlich_NPA,Marty,
G_CAPSS,Weise2,Kinkar,G_IFT,Deb,Tawfik,G_PQM} gave the estimation of $\eta$ in both hadronic
and quark temperature domain. There are 
also results, based on numerical simulation, addressed in Refs \cite{Bass,Muronga,Plumari,Pal}.
%
%The rising behavior of $\eta$ with T is shown in Refs\cite{Itakura,Dobado,Weise2}.
 
 % % % % % % % % % % % % % % % % % % % % % % %
% electrical conductivity

In this work we also estimate the 
electrical conductivity $\sigma$ of the hadronic system which  
has rich phenomenological and theoretical implications in heavy ion physics. 
The electrical conductivity can 
be associated to the rate of soft dilepton production \cite{Moore}
and photon multiplicity near zero transverse momentum~\cite{Nicola,Nicola_PRD}.
It also controls the rate of decay of the magnetic 
fields in the system produced in heavy ion collisions~\cite{Tuchin}.

Several authors~\cite{Marty,Puglisi,Greif,Cassing,TSteinert,Finazzo,SQin,LQCD_Ding,
LQCD_Arts_2007,LQCD_Buividovich,LQCD_Burnier,LQCD_Gupta,LQCD_Barndt,LQCD_Amato,
PKS,G_IFT2,Nicola_PRD,Greif2,G_CU,SS_rho,Lee} 
have calculated $\sigma$ by employing different methods  such as  -
unitarization of chiral perturbation theory~\cite{Nicola_PRD,Nicola}, numerical solution of  Boltzmann 
transport equation~\cite{Puglisi,Greif}, dynamical quasi-particle model~\cite{Marty}, 
off-shell transport model \cite{Cassing,TSteinert}, techniques of 
holography \cite{Finazzo},
Dyson - Schwinger approach \cite{SQin} and lattice gauge theory \cite{LQCD_Ding,LQCD_Arts_2007,LQCD_Buividovich,
LQCD_Burnier,LQCD_Gupta,LQCD_Barndt,LQCD_Amato}. 

Among the earlier works  
we  find that Refs.~\cite{Itakura,Dobado,Nicola,Weise1,SSS,Ghosh_piN,Gorenstein,HM,Hostler,Khvorostukhin}
and Refs.~\cite{Nicola_PRD,Greif2,G_CU,SS_rho,Lee} deal with shear and electrical conductivity of hadronic
matter respectively  where
majority of these in~\cite{Itakura,Dobado,Nicola,Weise1,SSS,Ghosh_piN,Nicola_PRD,G_CU,SS_rho,Lee} 
considered hadrons in the $u$, $d$ sector only. 
Shear viscosity has been estimated by including strangeness in hadron resonance gas (HRG) model 
in Refs~\cite{Gorenstein,HM,Hostler,Greif2}.
Though some of the effective QCD model calculations,
mentioned above, also included strange quark sector but their  contribution
has not been discussed separately. Since the melting of strange quark condensate is quite different
from condensate of u and d quarks, so the contribution of the strange sector in transport
coefficients may be necessary to consider separately. This fact may be supported from the 
recent Ref.~\cite{PNJL_V}, where a second peak of bulk viscosity 
is observed because of the strange quark contribution.

Calculating the relaxation time of kaon,
its contribution to shear viscosity of hadronic matter 
has been discussed in Refs~\cite{Kapusta:2008vb,Prakash}.
Here we are interested to focus on this issue with the help
of effective hadronic interactions in the strange sector.
In Refs.~\cite{Itakura,Nicola,SSS,Ghosh_piN,Nicola_PRD,G_CU,SS_rho}, 
we have seen that effective interactions
of pion with $\sigma$ and $\rho$ resonances are quite successful to
describe the (nearly) perfect fluid nature of hadronic matter. 
In the strange sector, one can use the effective interaction
of kaon with $K^*$ and $\phi$ resonances, whose contribution in
shear viscosity and electrical conductivity is the aim of present investigation.
%
%
%These literature covers a wide range of numerical 
%values of $\sigma$. Among them Refs \cite{Cassing,Marty,Nicola_PRD,Nicola,Greif2,G_CU} have observed 
%the decreasing nature of $\sigma(T)/T$ in temperature domain relevant for hadronic phase and increasing 
%trend in the temperature 
%domain relevant for QGP phase~\cite{Marty,Puglisi,Greif,Cassing,Finazzo,LQCD_Amato,PKS}. 
%However, it has   shown in the Refs.~\cite{Finazzo,Lee,G_IFT2} that
%$\sigma/T$ increases with $T$ in hadronic phase.
%The Uncertainty appears not only in   nature but also in their numerical values, 
%with a range $\sigma/T\approx 10^{-3}$ to $10^{-2}$ for hadronic phase.
%The estimates  from earlier investigations on transport coefficients of hadronic matter do not converge 
%which pave way for further investigation.
%
%
%In this work  we have mainly 
%focused on the contribution of K meson component to shear viscosity and
%electrical conductivity of hadronic matter. 
These transport coefficients are estimated by using their standard expressions, obtained from 
relaxation time approximation (RTA) or Kubo formalism with one-loop diagram, 
where the relaxation time or thermal width has been obtained from the effective
hadronic Lagrangian density of strange sector. 
%Owing to the optical theorem
%of thermal field theory, we have calculated kaon thermal width from the imaginary
%part of kaon self-energy for different possible mesonic loops.

The paper is organized as follows. In the next section, we have addressed the formalism 
to calculate the shear viscosity and electrical conductivity of kaon component 
in term of its thermal width. This section also includes the derivation of kaon thermal width from
different possible loop and scattering diagrams. The numerical results on these transport coefficients for kaon
component have been discussed in  section III. Section IV is devoted to  summary and conclusion.
%
%%%%%%%%%%%%%%%%%%%%%%%%%%%%%%%%%%%%%%%%%%%%%%%%%%%%%%%%%%%%%%%%%%%%%%%%%%%
\section{Formalism}
\label{sec:form}
%\vspace{0.5cm}
%
\begin{figure}
\begin{center}
\includegraphics[scale=0.5]{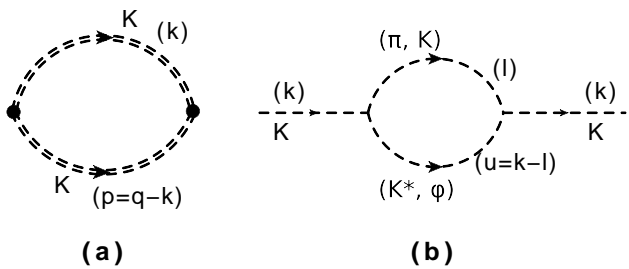}
\caption{The diagram (a) is a schematic one-loop representation of viscous-stress
tensor for the medium with constituents of K-meson. The double dashed lines
for the K-meson propagators indicate that they have some finite thermal
width. The diagram (b) is kaon self-energy for $\pi K^{*}$ and $K \phi$ loops.} 
\label{transport_K}
\end{center}
\end{figure}
\begin{figure}
\begin{center}
\includegraphics[scale=0.35]{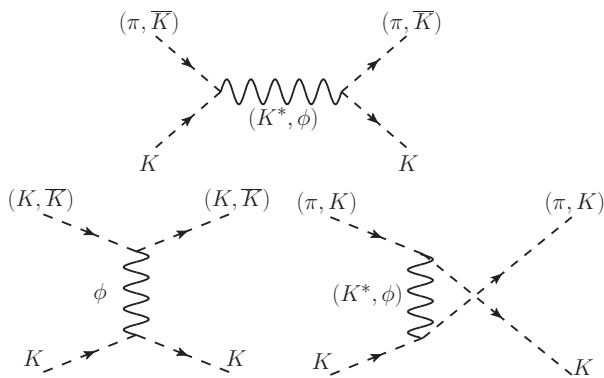}
\caption{Tree-level Feynman diagrams of $Kx\rightarrow Kx$ scattering, where $x=K,{\ov K}, \pi$.} 
\label{ScDiagram_K}
\end{center}
\end{figure}
We consider a hot mesonic matter, where pion and 
kaon are the main constituents of the medium. 
Here, our main focus is on the constituent of strange 
sector meson - the kaons. In the next section we derive the
expression for viscosity in terms of the width of the kaons
in thermal bath.

\subsection{Shear viscosity}
We start with
the viscous-stress tensor $\pi^{\mn}$ for kaonic medium
to construct the spectral density 
\be
A^{\eta}_K=\int d^4x e^{iq\cdot x}\langle[\pi_{ij}(x),\pi^{ij}(0)]\rangle_\beta~,
\label{spec_eta}
\ee
where $\langle \hat{O}\rangle_\beta$ stands for the thermal ensemble average of  $\hat{O}$ {\it i.e.}
$\langle \hat{O}\rangle_\beta={\rm Tr}{e^{-\beta H}\hat{O}}/{{\rm Tr}e^{-\beta H}}$.
With the help of the Kubo formalism~\cite{Zubarev,Kubo},
one can relate this spectral density $A^{\eta}_K$ with 
the shear viscosity ($\eta_K$) of kaonic medium as~\cite{Nicola} 
\be
\eta=\frac{1}{20}\lim_{q_0,\vq \rightarrow 0}\frac{A^{\eta}_K}{q_0}~.
\label{eta_Nicola}
\ee
Following quasi-particle method of Kubo 
framework~\cite{Nicola,Weise1,Sarkar_Book,G_Kubo}, 
the simplest one-loop expressions of Eq.~(\ref{eta_Nicola}) 
for kaon ($K$) is: 
\be
\eta_K=\frac{\beta I_K}{30\pi^2}\int^{\infty}_{0} 
\frac{d\vk\vk^6}{{\om^K_k}^2\Gamma_K}n_k(\om^K_k)
\{1+n_k(\om_k^K)\}~, 
\label{eta_K}
\ee
where isospin factor $I_K=4$ and $n_k(\om^K_k)=1/\{e^{\beta\om^K_k}-1\}$ is the
Bose-Einstein (BE) distribution of kaon with energy 
$\om^K_k=(\vk^2+m_K^2)^{1/2}$. 
Fig.~\ref{transport_K}(a) represents the schematic one-loop diagram,
derived from two point function of viscous stress tensor in 
the kaonic medium. $\Gamma_K$ in Eq.~(\ref{eta_K}) is the 
thermal width of kaon in the medium which is represented 
in  Fig.~\ref{transport_K}(a) by double lines. Hence, these internal lines are drawn in 
double line pattern.
This adoption of finite thermal width is a very 
well established technique~\cite{Nicola,Weise1,Sarkar_Book,G_Kubo},
which is generally used in Kubo approach
to get a non-divergent value of the shear viscous coefficient.
This treatment is equivalent to quasi-particle approximation.

Again, this one-loop expression of shear viscosity
from Kubo approach~\cite{Nicola,Weise1,Sarkar_Book,G_Kubo}
exactly coincides with the expression originating from the relaxation-time 
approximation of the kinetic theory approach~\cite{SSS,Sarkar_Book,Gavin,Prakash}
%---------After report2 from referee------------------------
Although, RTA is used in the present
work for simplicity, we should, however,  mention here 
that the re-summation of ladder diagrams 
~\cite{Nicola,Jeon1,Jeon2,Jeon3,Arnold1,Arnold2,Arnold3} leads to the  
contribution to shear viscosity which has the 
same order of magnitude as the one loop diagram.
%------------------------------------------------------------
The dissipative part of energy
momentum tensor, responsible for shear viscosity coefficient, can be
expressed as
\bea
T^{\mu\nu}_D&=&\eta (D^\mu u^\nu + D^\nu u^\mu +\frac{2}{3}
\Delta^{\mu\nu}\partial_\sigma u^\sigma)
\nn\\
&=&\int \frac{d^3\vk}{(2\pi)^3}\frac{k^\mu k^\nu}{\om^K_k}\delta n~,
\label{RTA_Tmn}
\eea
where 
\be
D^\mu=\partial^\mu - u^\mu u_\sigma \partial_\sigma,~
{\rm and}~\Delta^{\mu\nu}=u^\mu u^\nu - g^{\mu\nu}~.
\ee
It is interesting to note that Eq.~(\ref{RTA_Tmn}) relates the collective fluid  
four velocity,  $u^\mu$ to the four momentum of elementary constituent,  $k^\mu$ through
the non-equilibrium distribution function $n=n_k + \delta n$, which is assumed to be
slightly shifted from equilibrium distribution function $n_k$ by 
$\delta n$,  given by 
\be
\delta n =C k_\mu k_\nu(D^\mu u^\nu + D^\nu u^\mu +\frac{2}{3}
\Delta^{\mu\nu}\partial_\sigma u^\sigma)n_k (1+n_k)~,
\label{delta_n}
\ee
The Boltzmann equation in RTA can be written as,
\be
k^\mu \partial_\mu n_k =\frac{\om^K_k}{\tau_K}\delta n~,
\ee
which fixes the value of $C$ as $C=\frac{\tau_K\beta}{2\om^K_k}$.
Using this expression of $C$ in Eq.~(\ref{delta_n})  
one can obtain Eq.~(\ref{eta_K}) through Eq.~(\ref{RTA_Tmn})
in RTA approach where we have to
identify the kaon relaxation time $\tau_K=1/\Gamma_K$.
Hence, the thermal width or relaxation time of kaon plays a vital
role in determining the  numerical strength of shear viscosity. 
%for kaon component. 
To calculate this kaon relaxation time, we have considered 
two sources of Feynman diagrams. One is loop diagram of kaon self-energy 
and another is tree-level diagram for elastic scattering channel of kaon. 
These  are described below.

\subsection{Loop diagrams}
\label{LoopD}
Following the Cutkosky rule, we will estimate the thermal width from the 
imaginary part of kaon self-energy at finite temperature. Here, we have
considered $\pi K^*$ and $K \phi$ loops for calculating kaon self-energy,
which is shown in Fig.~\ref{transport_K}(b).
We can write the total thermal width/Landau damping (LD) of kaon $\Gamma_K$ as
\bea
\Gamma^{LD}_K&=&-{\rm Im}{\Pi}^{LD}_{K(\pi K^*)}(k_0=\om^K_k,\vk)/(2\om_K)
\nn\\
&&-{\rm Im}{\Pi}^{LD}_{K(K \phi)}(k_0=\om^K_k,\vk)/(2\om_K)~,
\nn\\
\label{Gam_K}
\eea
where ${\Pi}^{LD}_{K(\pi K^*)}(k)$ and ${\Pi}^{LD}_{K(K \phi)}(k)$ 
are kaon self-energy for $\pi K^*$ and $K \phi$ loops respectively.
%The superscripts $R$ stands for retarded component of self-energy and 
The subscript,
$K$ indicates the external mesons and the mesons within parenthesis stand for 
those present in the internal lines of the kaon
self-energy graphs as shown in Fig.~\ref{transport_K}(b).

The imaginary part of kaon self-energy 
for $\pi K^*$ and $K \phi$ loops are respectively given as:
\bea
{\rm Im}{\Pi}^{LD}_{K(\pi K^*)}(k) &=&  \int \frac{d^3 \vl}{32\pi^2 \om^{\pi}_l\om^{K^*}_u} 
L_{K(\pi K^*)}\left(k,l \right)|_{l_0=-\om^{\pi}_l}
\nn\\
&& 
\{n_l(\om^\pi_l) - n_u(\om^{K^*}_u=k_0 + \om^\pi_l)\}
\nn\\
&&~~\delta(k_0 +\om^{\pi}_l - \om^{K^*}_u)~,
\label{G_KpiKst}
\eea
and
\bea
{\rm Im}{\Pi}^{LD}_{K(K\phi)}(k) &=& \int \frac{d^3 \vl}{32\pi^2 \om^{K}_l\om^{\phi}_u} 
L_{K(K\phi)}\left(k,l \right)|_{l_0=-\om^{K}_l}
\nn\\
&& 
\{n_l(\om^K_l) - n_u(\om^{\phi}_u=k_0 + \om^K_l)\}
\nn\\
&&~~\delta(k_0 +\om^{K}_l - \om^{\phi}_u)~,
\label{G_KKfi}
\eea
where $n_l(\om^\pi_l)$, $n_u(\om^{K^*}_u=k_0 + \om^\pi_l)$, 
$n_l(\om^K_l)$, $n_u(\om^{\phi}_u=k_0 + \om^K_l)$, 
are BE distribution functions of $\pi$, $K^*$, $K$ and $\phi$
mesons respectively.
%and the limits of integration are
%\bea
%\om^\pi_{l\pm} &=& \frac{(k^2 +m_\pi^2-m_{K^*}^2)}{2k^2} 
%\left[- k_0 \pm |\vk| \, \{1
%\right.\nn\\
%&&\left.- {4m_\pi^2k^2}/{(k^2 
%+m_\pi^2-m_{K^*}^2)^2}\}^{1/2} \right] ,
%\eea
%
%and 
%\bea
%\om^K_{l\pm} &=& \frac{(k^2 +m_K^2-m_{\phi}^2)}{2k^2} 
%\left[- k_0 \pm |\vk| \, \{1
%\right.\nn\\
%&&\left.- {4m_K^2k^2}/{(k^2 
%+m_K^2-m_{\phi}^2)^2}\}^{1/2} \right] ~.
%\eea

Using the interaction Lagrangian densities~\cite{CM_Ko} :
\be
{\cal L}^{\rm int}_{K\pi K^*} = ig_{K\pi K^*}[{\ov K^{*\mu}}\cdot{\vec \tau}
\{K(\del_\mu \pi) - ({\del_\mu K})\pi\}] + h.c.~,
\label{Lag_KpiKs}
\ee
and 
\be
{\cal L}^{\rm int}_{KK\phi}= g_{KK\phi}[\phi^\mu\{{\ov K}(\del_\mu K) - ({\del_\mu {\ov K}}) K \}] ~,
\label{Lag_KKphi}
\ee
we obtain the $K\pi K^*$ and $KK\phi$ vertices as:
\bea
L(k,l)_{K\pi K^*}&=& g_{K\pi K^*}^2[\{k^2 + l^2 +2 (k\cdot l)\}  
\nn\\
&& -\frac{(k^2 - l^2)^2}{m_{K^*}^2}\big ]~,~
{\rm for}~\pi K^*~{\rm loop}~,
\label{L_KpiKs}
\eea
and 
\bea
L(k,l)_{KK\phi}&=&g_{KK\phi}^2[\{k^2 + l^2 +2 (k\cdot l)\}  
\nn\\
&& -\frac{(k^2 - l^2)^2}{m_{\phi}^2}\big ]~,~
{\rm for}~K\phi~{\rm loop}~.
\eea
The effective coupling constants $g_{K\pi K^*}/(4\pi)=0.86$
and $g_{KK\phi}/(4\pi)=1.82$ are fixed~\cite{CM_Ko} from the experimental
decay widths of the processes $K^*\rightarrow K\pi$ and $\phi\rightarrow K{\ov K}$
respectively.

\subsection{Elastic scattering diagrams}
\label{ScD}
The $2\leftrightarrow 2$ kind of elastic scattering diagrams, shown in Fig~(\ref{ScDiagram_K}), are another
possible source along with the LD source, which basically represents $1\leftrightarrow 2$ type inelastic
processes in the medium~\cite{Weldon}. The expression of $\Gamma_K^{Sc}$ for $Kx\rightarrow Kx$
scatterings is
\be  
\Gamma_K^{Sc}=\sum_{x=K,{\ov K},\pi}\int \frac{d^3\vp}{(2\pi)^3}v_{Kx}\sigma_{Kx}n_p^x(1+n_p^x)~,
\label{GK_Sc}
\ee
where relative velocity between $K$ and $x$ meson is given by,
\bea
v_{Kx}&=&\frac{[\{s-(m_K +m_x)^2\}\{s-(m_K - m_x)^2\}]^{1/2}}{2\om_K\om_x}~,
\nn\\
{\rm with}~s&=&(\om_K+\om_x)^2,~\om_K=\{\vk^2+m^2_K\}^{1/2},
\nn\\
\om_x&=&\{\vp^2+m^2_x\}^{1/2},
\eea
and $n^x_p=1/\{e^{\beta\om_x}-1\}$ is the BE distribution 
function for $x$ meson. 
From different possible tree-level elastic scattering diagrams,
shown in Fig.~(\ref{ScDiagram_K}), the cross section $\sigma_{Kx}$ of $Kx$ scattering
are obtained as
\begin{equation}
\sigma_{Kx} =\frac{1}{16 \pi \lambda(s,m_K^2,m_x^2)}\int\limits_{t_{min}}^{t^{max}}|\ov {M_{Kx}}|^2 dt~,
\end{equation}
where $\lambda(s,m_K^2,m_x^2)=s^2+m_K^4+m_x^4-2sm_K^2-2sm_x^2-2m_K^2m_x^2$, known 
as the Källén  function and
%\begin{equation}
%\frac{d\sigma }{dt}=\frac{|\ov {M_{Kx}}|^2}{16 \pi \lambda(s,m_K^2,m_x^2)}~.
%\end{equation}
the limits of the above integration are $t_{max}=0$ and $t_{min}=-\lambda(s,m_K^2,m_x^2)/s$.
With the help of the same interaction Lagrangian densities,
given in Eqs.~(\ref{Lag_KpiKs}) and (\ref{Lag_KKphi}), we can construct
matrix elements of three different elastic channels - (1). $KK\rightarrow KK$, 
(2). $K\pi\rightarrow K\pi$,
and (3). $K{\ov K}\rightarrow K{\ov K}$, 
which are respectively written below. The 
square of the iso-spin averaged amplitude 
for the elastic scattering of particles, $K(k)$  and $x(p)$ with incident momenta
$k$ and $p$ respectively  is generically denoted by $|\ov{M_{Kx}}|^2$. 
The relevant amplitudes are as follows: 
(1)  The square of the iso-spin averaged amplitude
for $K(k)K(p)\longrightarrow K(k')K(p')$ is:
\begin{equation}
|\ov{M_{KK}}|^2
%=\frac{\sum\limits_{I} (2I+1)|M_I|^2}{\sum\limits_{I} (2I+1)}
=\frac{1}{4} \Big [3|M^1_{KK}|^2+|M^0_{KK}|^2 \Big]~,
\end{equation}
where
\begin{equation}
M^1_{KK}=g_{\phi KK}^2  \Big [\frac{(u-s) }{(t-m_\phi ^2 +i\epsilon)}+\frac{(t-s) }{(u-m_\phi^2+i\epsilon) } \Big]~,
\end{equation}
\begin{equation}
M^0_{KK}=g_{\phi KK}^2  \Big [\frac{(u-s) }{(t-m_\phi ^2 +i\epsilon)}-\frac{(t-s) }{(u-m_\phi^2+i\epsilon) }\Big]~.
\end{equation}
(2) Amplitude for $K(k)\pi (p)\longrightarrow K(k')\pi (p')$ is,
\begin{equation}
|\ov{{M_{K\pi}}}|^2
%=\frac{\sum\limits_{I} (2I+1)|M_I|^2}{\sum\limits_{I} (2I+1)}
=\frac{1}{3} \Big [2|M^1_{K\pi}|^2+|M^0_{K\pi}|^2 \Big]~,
\end{equation}
where
\begin{equation}
M^{3/2}_{K\pi}=2g_{\pi KK^*}^2  \Big [\frac{(t-s)+(m_K^2-m_\pi ^2)^2/ m_{K^*}^2 }{(u-m_{K^*}^2 
+i\epsilon)} \Big]~,
\end{equation}
\bea
M^{1/2}_{K\pi}&=&g_{\pi KK^*}^2  \Big [3\{\frac{(t-u)+(m_K^2-m_\pi ^2)^2/ m_{K^*}^2 }{(s-m_{K^*}^2 +i\epsilon)}\} 
\nn\\
&&-\{\frac{(t-s)+(m_K^2-m_\pi ^2)^2/ m_{K^*}^2 }{(u-m_{K^*}^2 +i\epsilon)}\}\Big]~.
\eea
(3) Similarly for the process, $K(k)\ov{K}(p)\longrightarrow K(k')\ov{K}(p')$ the amplitude is:
\begin{equation}
|\ov{M_{K{\ov K}}}|^2
%=\frac{\sum\limits_{I} (2I+1)|M_I|^2}{\sum\limits_{I} (2I+1)}
=\frac{1}{4}\Big [3|M^1_{K{\ov K}}|^2+|M^0_{K{\ov K}}|^2 \Big]~,
\end{equation}
where
\begin{equation}
M^1_{K{\ov K}}=g_{\phi KK}^2  \Big [\frac{ s-u  }{ t-m_\phi ^2 +i\epsilon }\Big]~,
\end{equation}
\begin{equation}
M^0_{K{\ov K}}=g_{\phi KK}^2 \Big [\frac{2(t-u) }{(s-m_\phi ^2 +i\epsilon)}+\frac{(s-u) }{(t-m_\phi^2+i\epsilon) }\Big]~.
\end{equation} 
In the above expressions, the Mandelstam variables are $s=(k+p)^2$, $t=(k-k')^2$, $u=(k'-p)^2$.
We have also taken the experimental values of decay widths for $K^*$ and $\phi$ mesons propagators
in $s$-channel diagrams of Fig.~(\ref{ScDiagram_K}).
%--------------------------------
\subsection{Nucleonic matter}
\label{NuclMat}
Now  we would like to estimate the effects of kaon-nucleon interaction  
on the width of kaons in the thermal bath. For this purpose the
natural way to proceed is to calculate the possible
baryon loops contributing to the kaon self-energy. The $N\Lambda$ and $N\Sigma$
can be considered as possible candidates.
The Landau and unitary cut contributions 
to kaon self-energy can be calculated using 
effective $KN\Lambda$ and $KN\Sigma$ interaction Lagrangian densities.
However, we do not get any on-shell value for kaon relaxation time
because kaon pole ($m_k$) is located neither in its Landau
cut region ($0$ to $|m_{\Lambda,\Sigma}-m_N|$) nor in its unitary
cut region ($|m_{\Lambda,\Sigma}+m_N|$ to $\infty$), where $m_N=0.940$
GeV, $m_\Lambda=1.115$ GeV and $m_\Sigma=1.189$ GeV are the masses of
$N$, $\Lambda$ and $\Sigma$ baryons respectively. Therefore, instead of
this methodology of loop calculation, we have resorted to the following alternative  
way.

Let us take the experimental values of scattering length $a_{KN}^I$ of $KN$
interaction, where $I$ stands for different isospin states of $KN$ system.
From Refs.~\cite{exp1,exp2,exp3}, using the scattering lengths 
$a_{KN}^{I=0}=-0.007$ fm and $a_{KN}^{I=1}=-0.225$ fm, we obtain
the isospin averaged cross section,
\be
\sigma_{KN}=4\pi \sum_{I=0,1} (2I+1)|a_{KN}^{I}|^2\Big/\sum_{I=0,1}(2I+1)
\approx ~4.7~ {\rm mb}~,
\ee
which can be used to calculate the relaxation time $\tau_{KN}$ or thermal width $\Gamma_{KN}$ 
of $K$ by using Eq.~(\ref{GK_Sc}), where $x$ meson is replaced by nucleon $N$
and its distribution function will be (Fermi-Dirac) $n^N_p=1/\{e^{\beta\om_N}+1\}$. 
%
%\be
%\Gamma_K=1/\tau_{K}=\int \frac{d^3\vl}{(2\pi)^3} ~\sigma_{K N}~v_{K N}~n_{N} ~,
%\label{G_KN}
%\ee
%where relative velocity of $KN$ system is given by,
%\bea
%v_{KN}&=&\frac{[\{s-(m_K +m_N)^2\}\{s-(m_K - m_N)^2\}]^{1/2}}{2\om_K\om_N}~,
%\nn\\
%{\rm with}~s&=&(\om_K+\om_N)^2,~\om_K=\{\vk^2+m_K\}^{1/2},
%\nn\\
%\om_N&=&\{\vl^2+m_N\}^{1/2},
%\eea
%and $n_N=1/\{e^{\beta(\om_N -\mu_N)}+1\}$ is the Fermi-Dirac distribution 
%function for nucleon.
%

Finally, for total thermal width of $K$, we have to add LD and scattering part
of mesonic matter, described in subsection (\ref{LoopD}), (\ref{ScD}) and the
nucleonic matter contribution, described in subsection (\ref{NuclMat}).
We however find that the nucleonic matter has negligibly small numerical contribution.

%--------------------------------------------

\subsection{Electrical conductivity}
Similar to shear viscosity, one can derive the expression for 
electrical conductivity,  $\sigma_K$ for kaon component,
using the spectral density of current-current correlator
\be
A^{\sigma}_K=\int d^4x e^{iq\cdot x}\langle[J_{i}(x), J^{i}(0)]\rangle_\beta~.
\label{spec_cond}
\ee
The corresponding Kubo relation reads,
\be
\sigma_K=\frac{1}{6}\lim_{q_0,\vq \rightarrow 0}\frac{A^{\sigma}_K}{q_0}~.
\label{cond_Nicola}
\ee
Within the one loop approximation the expression for $\sigma_K$ is given by~\cite{Nicola_PRD,G_CU}
\be
\sigma_{K}=\frac{g^e_{K}}{3T}\int \frac{d^3\vk}{(2\pi)^3\Gamma_K}
\left(\frac{\vk}{\om^k_{K}}\right)^2
\left[ n_{K}(\om^k_K)\{1 + n_{K}(\om^k_K)\}\right]~,
\label{el_K}
\ee
where $g^e_{K}=2 e^2$ is isospin factor for charged kaon - $K^+$ and $K^-$.
Same expression  for  $\sigma_K$ can be obtained in RTA approach.
% 
%
% 
%%%%%%%%%%%%%%%%%%%%%%%%%%%%%%%%%%%%%%%%%%%%%%%%%%%%%%%%%%%%%%%%%%%%%%%%%%%
\section{Results and Discussion}
\label{sec:num}
\begin{figure}
\begin{center}
\includegraphics[scale =0.35]{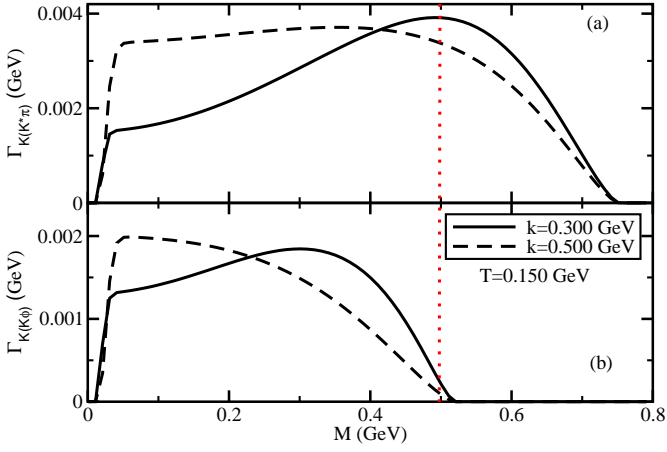}
\caption{(Color online) The thermal width of kaon self-energy from (a)$K^{*}\phi$
and (b) $K\phi$ loops as a function of invariant mass 
${M=\sqrt{k_{0}^{2}-\rvert{\bold{k}\lvert^2}}}$  at fixed temperature 
$T=0.150$ GeV for two different momenta $\vk =0.300$ GeV (solid line) 
and $0.500$ GeV (dashed line). 
The vertical red dotted line represents the on-shell value $M=m_{K}$.}
\label{GK_M}
\end{center}
\end{figure}
\begin{figure}
\begin{center}
\includegraphics[scale=0.35]{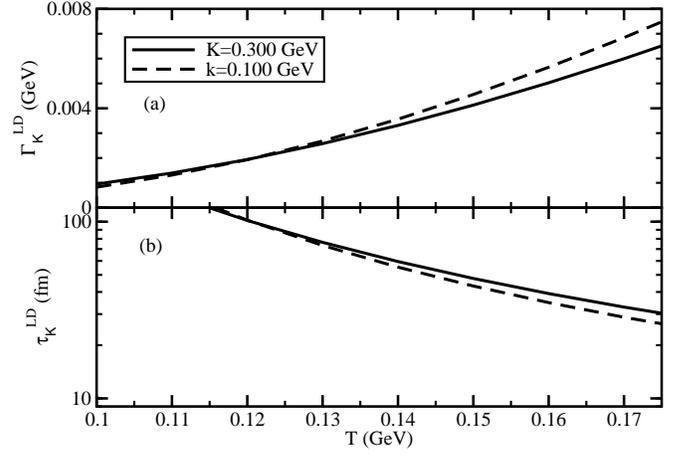}
\caption{ Temperature dependence of  (a) thermal width and (b) the mean 
free path of kaon for two different momenta $\vk$=0.100 GeV (solid line)
and $\vk$=0.300 GeV (dashed line). 
The horizontal dashed red line represents an approximate dimension of fireball.} 
\label{GK_T_k}
\end{center}
\end{figure}
\begin{figure}
\begin{center}
\includegraphics[scale=0.35]{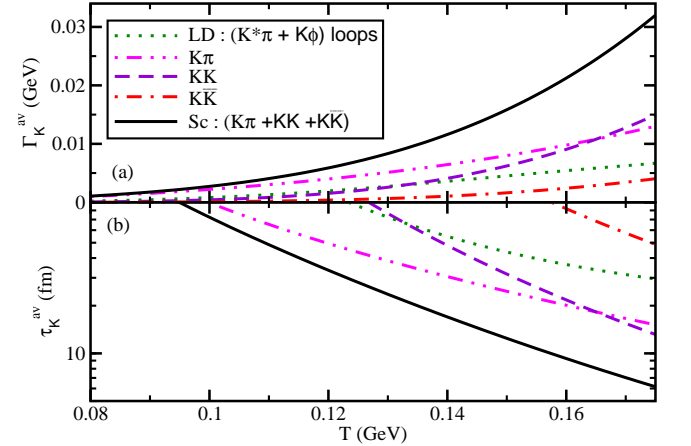}
\caption{(Color online) $T$ dependence of average thermal widths (a) and relaxation times (b) 
for $KK$ (dashed line), $K\pi$ (dash-double-dotted line), $K{\ov {K}}$ (dash-dotted line)
scatterings and their total (solid line) and Landau damping part (dotted line).} 
\label{GLS_T}
\end{center}
\end{figure}
First we explore the contribution of loop diagram in kaon relaxation time.
From Eqs.~(\ref{G_KpiKst}) and (\ref{G_KKfi}), one can respectively obtain 
the individual contributions from $K^*\pi$ and $K\phi$ loops 
to the (off-shell) thermal width of kaon as a function of the invariant 
mass ${M=\sqrt{k_{0}^{2}-\rvert{\bold{k}\lvert^2}}}$ for two different values of three 
momenta $\bold{k}=0.300 $ GeV and $\bold{k}=0.500 $ GeV at temperature $T=0.150$ GeV. 
The results are shown in Fig.~(\ref{GK_M}), where we have used $m_\pi=0.140$ GeV, $m_K=0.500$ GeV, 
$m_{K^*}$=0.890 GeV, $m_\phi=1.020$ GeV.		 
It is clear that the Landau cuts~\cite{Weldon,Ghosh_piN} end sharply at $M=|m_\phi -m_K|=0.51$ GeV for the $K\phi$ 
loop and $M=|m_{K^*}-m_\pi|=0.745$ GeV for $K^*\pi$ loop. 
%At lower invariant mass, the thermal width of the 
%two loops are much larger for higher three momentum while 
%the tail parts show no such momentum dependence.(why ?)
The red vertical dashed line in Fig.~(\ref{GK_M}) denotes the physical pole mass
of kaon and its corresponding contribution will provide on-shell values
of kaon thermal widths for $K^*\pi$ and $K\phi$ loops respectively. One should
keep in mind that these Landau cut contributions only originate
in the presence of medium. In vacuum, we can not have any (on-shell) decay like
$K\rightarrow K^*\pi$ or $K\rightarrow K\phi$ because of kinematic restrictions. 
Here we get a non-zero thermal width of kaon because of in-medium
$KK^*\pi$ and $KK\phi$ scatterings which were absent in vacuum.

In Fig.~\ref{GK_T_k} we display the temperature dependence of total (on-shell) 
thermal width $\Gamma^{LD}_K$ and relaxation time $\tau^{LD}_K=1/\Gamma^{LD}_K$ of kaon for 
$\vk=0.100$ GeV (solid line) and $\vk=0.300$ GeV (dashed line).
The total (on-shell) thermal width of kaon is the summation of
$K^*\pi$ and $K\phi$ loop contributions. We observe that 
the contributions from the former dominates over the latter
as revealed in Fig.~(\ref{GK_M}). 
We observe that the thermal width increases with the temperature.
% %is nearly insensitive to  the three momentum 
As relaxation time $\tau_K^{\rm LD}$ is inversely proportional to the thermal width ($\Gamma_K$), the 
$\tau^{LD}_K$ decreases rapidly as temperature increases. 
%We also notice that the 
%kaon thermal width is not too sensitive to the three momenta as opposed to the mean free path
%which is strongly affected by the change in momentum. 
%----------
We have seen that the nucleonic contribution due to $KN$ interaction
to the kaon thermal width or mean free path 
is insignificant compared to the contributions from mesonic loops, 
therefore, the nucleonic part has not been shown separately.
%----------
%The horizontal red dashed line indicates an approximate dimension
%of fireball, produced in heavy ion experiments like RHIC and LHC. 
%One can notice that
%dashed and solid lines in Fig.~\ref{GK_T_k}(b) have different temperature ranges, where
%their mean free paths remain within the fireball dimension. We expect that kaon dissipation
%within these temperature ranges will  contribute to the  transport coefficients
%of the fireball. The kaon, having mean free path greater than this fireball dimension, 
%will leave the system without facing any interaction and hence will not contribute to 
%the transport coefficients.
%
\begin{figure}
\begin{center}
\includegraphics[scale=0.35]{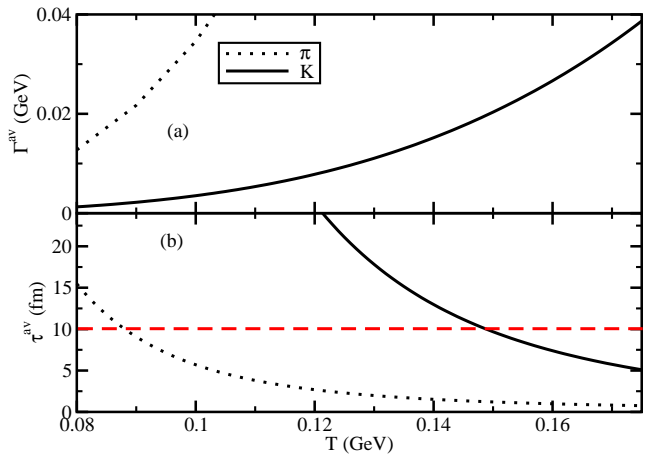}
\caption{(Color online) Temperature dependence of average (a) thermal widths and (b) 
relaxation times of pion (dotted lines) and kaon (solid lines) 
The horizontal dashed line represents an approximate life time of fireball.} 
\label{Gav_T}
\end{center}
\end{figure}
\begin{figure}
\begin{center}
\includegraphics[scale=0.35]{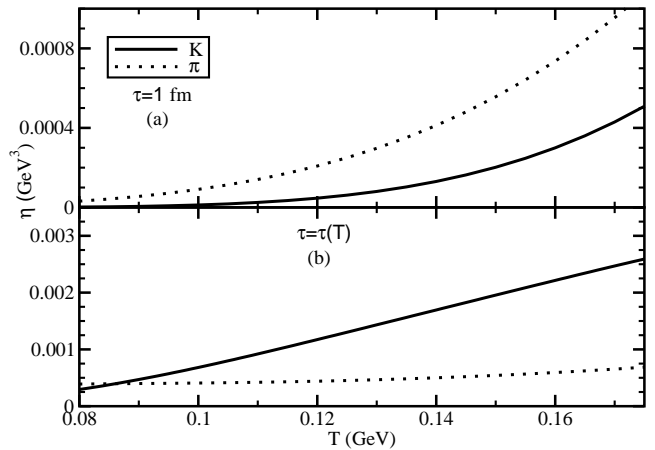}
\caption{Temperature dependence of shear viscosity for 
kaon (solid line) and pion (dotted line) by using
(a) $\tau =1$ fm and (b) $\tau=\tau (T)$.}
%In (b), dashed
%line represents shear viscosity Kaon without taking LD part
%and dash-dotted line gives the addition of pion and kaon viscosities.} 
\label{eta_T}
\end{center}
\end{figure}
\begin{figure}
\begin{center}
\includegraphics[scale=0.35]{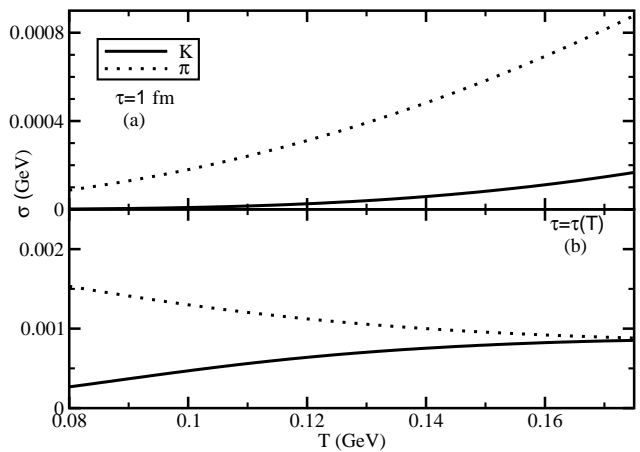}
\caption{Same as Fig.~\ref{eta_T} for electrical conductivity.} 
\label{el_T}
\end{center}
\end{figure}
We can take thermal average of $\Gamma^{LD}_K(\vk,T)$ by using the relation
\be
\Gamma^{\rm av}_K=\int_0^\infty\frac{d^3\vk}{(2\pi)^3} \Gamma^{LD}_K(\vk, T) n^K_k(\om^K_k)
{\Big/}\int_0^\infty\frac{d^3\vk}{(2\pi)^3} n^K_k(\om^K_k)~,
\label{G_av}
\ee
which is plotted using dashed line in Fig.~\ref{GLS_T}(a). Similar to LD source, different
scattering contributions for $K\pi$, $KK$, $K{\ov {K}}$ interactions and their total are shown by
dash-double-dotted, dashed, dash-dotted and solid lines in Fig.~\ref{GLS_T}(a). Their corresponding
relaxation times are drawn in Fig.~\ref{GLS_T}(b). Here we notice that the contribution
of elastic scattering is dominating over the LD contribution. After adding them we get total
thermal width of kaon, which is plotted by solid line in Fig.~\ref{Gav_T}(a) and
it is also compared with the thermal width of pion component (dotted line), 
taken from Ref.~\cite{Ghosh_piN}. In Fig.~\ref{Gav_T}(b), their relaxation times are also 
compared and we noticed that the relaxation time of kaon is greater than that of pion.
We consider that the lifetime of fireball produced in heavy ion experiments is 
approximately $10$ fm as indicated by a horizontal (red) dashed line in Fig.~\ref{Gav_T}(b).
We observe that the relaxation time induced by the interactions involving 
pions is smaller than that involving kaons. The life time of the fireball 
is larger than the pions for almost the entire range of temperature considered.
However, the kaons  has the chance of getting equilibrated only at high 
temperatures, $T\sim 150$ MeV which is close to the quark-hadron transition temperature.

In Fig.~{\ref{eta_T}}, we present the temperature dependence of 
shear viscosity of kaon (solid lines) and compare with 
pion (dotted lines) for (a)  constant relaxation times and (b) temperature 
dependent relaxation times (displayed in Fig.~\ref{Gav_T}). 
%
%For constant relaxation time, shear viscosities of both 
%pion and kaon are very small at low temperature 
%but increase  as the temperature rises. 
For constant relaxation time, shear viscosity for kaon component is smaller 
than that of pion but for temperature dependent relaxation time the trend
is reversed. 

Eq.~\ref{eta_K} indicates that for a constant relaxation 
time (or thermal width) the temperature dependence of 
shear viscosity is governed  by the thermal phase space factors.
The massive kaons are thermally more suppressed than the
lighter pions in the bath, consequently the $\eta$ for kaons is smaller than
that of pions for the same value of relaxation time ($\tau=1$ fm/c). This is reflected 
through the results displayed in the upper panel of Fig.~\ref{eta_T}.  

The shear viscosity represents fluid resistance to the transfer of momentum
through collision processes. 
The momentum transfer gets easier and hence the fluid resistance or the shear 
viscosity becomes  smaller in a strongly interacting medium.  
Therefore, the inclusion of interactions of pions and kaons in evaluating their
shear viscosities is essential. We note that thermal 
width (the interaction rate) of kaons is smaller than pions since
the  interaction of kaons is weaker. 
%Therefore, in a more realistic scenario the viscosity depends 
%on the phase space as well on the dynamics (interaction) of the process. 
We find that the viscosity
of kaons is larger than the pions because  the thermal
suppression of kaons is overwhelmed by their larger 
relaxation time (smaller thermal width) as compared to pions. 
In other words the smaller relaxation time of pions
compensate the larger thermal abundance and hence
results in smaller viscosity. 
%-----------------------------------------------------
%Since the former case reflects the $T$ dependence of 
%phase space in $\eta$, so the massive kaons are  exposed to larger thermal
%suppression than the lighter pion. On the other hand, our calculated 
%relaxation time
%for kaon is little larger than that for pion, therefore, in that case, 
%the kaon component dominates over the pion component.
%-----------------------------------------------------
In the context of finite size RHIC or LHC matter the
kaon component becomes relevant at high $T$ because the kaon relaxation
time at $T>$150 MeV is smaller than than the typical fireball life time. 
Hence, the role of kaon viscosity near the
transition temperature cannot be ignored. 
Similar observation was 
made in earlier investigations~\cite{Kapusta:2008vb,Prakash} too.
%------------------------------
%In Fig.~\ref{eta_T}(b), dash line indicates shear viscosity
%of kaon component when thermal width of kaon is evaluated with the  mesonic loop only.
%The inclusion of nucleonic part to the mesonic loop leads to the 
%viscosity shown by solid line.  A mild suppression
%of shear viscosity at high temperature range is noticed and we realize 
%that the nucleonic part of Kaon thermal width is responsible for this fact.
%-----------------------------
%
%shear viscosity of kaon increases rapidly although for pion, it does not increases at 
%all and remains at very low value even at high temperature.
% 

Next, in Fig.~\ref{el_T}(a),
the temperature dependence of electrical conductivities for kaon (solid lines) and 
pion (dotted line) are drawn for constant relaxation time $\tau$ and similar to
Fig.~\ref{eta_T}(a), pion component is larger than kaon component due to the 
reason explained above - the  massive kaons are thermally suppressed more than pions. 

We find that with the temperature dependent thermal width (or relaxation time)
the electrical conductivity of  pion decreases 
unlike  the shear viscosity which shows a mildly increasing trend 
with  increase in temperature (Fig.~\ref{eta_T} b).
This difference originates from the expressions of $\sigma$ and $\eta$, given
by Eqs.~(\ref{el_K}) and (\ref{eta_K}) respectively involving different power of momentum 
in their integrand. This can be understood easily at the high temperature
limit ($T>>m_\pi$) where the electrical conductivity of pions, $\sigma_\pi^{el}\sim T^2/\Gamma^{av}(T)$
and the shear viscosity, $\eta_\pi\sim T^4/\Gamma^{av}(T)$ which clearly indicates that $\sigma_\pi^{el}$
and $\eta_\pi$ have different $T$ dependence.
For the kaon component, both $\eta_K(T)$ and $\sigma_K(T)$ are increasing functions but
their slopes are different because of the different exponent of momentum
appearing in the expressions for viscosity and electrical conductivity.
It is also interesting to note that the pion and kaon contributions
to the electrical conductivity is comparable at high $T$ and 
the nature of $T$ variation will be governed by their combined strengths.

%At constant relaxation time $(\tau =1$ fm), the figure (upper panel) shows that the electrical
%conductivity of kaon rises very slowly at low temperature but rises fast after T=0.14 GeV. The opposite picture 
%happens For pion in which the electrical conductivity slowly decreases and remains almost constant 
%as the  temperature rises.
%
%\begin{figure}
%\begin{center}
%\includegraphics[scale=0.35]{etaK_T_V.eps}
%\caption{(Color online) Temperature dependence of (a) threshold momentum or momentum cutoff and (b) 
%shear viscosity for kaon with momentum cutoff (dashed line), pion (dotted line) 
%and their total (solid line).} 
%\label{etaK_T_V}
%\end{center}
%\end{figure}
%
%
%\begin{figure}
%\begin{center}
%\includegraphics[scale=0.35]{el_T2.eps}
%\caption{Temperature dependence of (a) $\sigma$ and (b) $\sigma /T$ for kaon with 
%momentum cutoff (dashed lines), pion (dotted lines) and their total (solid lines).} 
%\label{el_T2}
%\end{center}
%\end{figure}
%  

%%%%%%%%%%%%%%%%%%%%%%%%%%%%%%%%%%%%%%%%%%%%%%%%%%%%%%%%%%%%%%%%%%%%%%%%%%%
\section{Summary and Conclusion}
In this work we have estimated the contributions of strange hadrons to 
shear viscosity and electrical conductivity of hadronic matter. It is well known
that the standard expressions of these transport coefficients can be obtained
either from RTA or Kubo framework, where the relaxation time of medium constituents
proportionally controls their numerical strengths. We have calculated kaon
relaxation time by considering different possible $1\leftrightarrow 2$ and
$2\leftrightarrow 2$ processes. Contributions from $1\leftrightarrow 2$
have  been calculated
from the imaginary part of kaon self-energy at finite temperature. For
the calculation of self energy $\pi K^*$ and 
$K\phi$ have been considered in the internal lines.
For $2\leftrightarrow 2$ type of elastic channels, $K\pi$, $KK$
and $K{\ov {K}}$ processes have been considered.  All these in-elastic and elastic channels are
calculated from an effective hadronic Lagrangian density, describing $KK\phi$ 
and $KK^*\pi$ interactions. We notice that $2\leftrightarrow 2$ scattering processes 
in medium dominate over the  $1\leftrightarrow 2$ processes. 
The inverse of the total scattering probability, quantified by total thermal width,
basically gives the relaxation time of kaons in the medium. If we consider
the life time of the hadronic fireball produced at RHIC and LHC collisions $\sim 10$ fm/c
then we notice that the kaon has the chance to attain equilibrium at the high 
temperature domain close to the transition temperature. 
It signifies that for phenomenological purpose, 
the contribution of kaon component is relevant in the high temperature region.

Using the total kaon relaxation time,
we have estimated shear viscosity and electrical conductivity of kaon component.
When we compare with the pion component, estimated in earlier work~\cite{Ghosh_piN}
using a similar type of effective hadronic interactions, we notice that
kaon has smaller electrical conductivity than pions but 
an opposite trend is observed in shear viscosity {\it i.e.} the viscosity of
kaons is larger than pions. 
%---------After report2 from referee----------------------------
%for almost the entire temperature range considered.
We  also note that with the inclusion of $K-\pi$ interaction the
magnitude of $\eta_K$  reduces compared to the scenario
when $K-\pi$ interaction is ignored in estimating $\eta_K$.
However, the inequality, $\eta_K>\eta_\pi$ is still maintained with $K-\pi$
interaction for almost the entire temperature range considered here
($\eta_\pi$ is shear viscosity of pion). 
That is the gap between the $\eta_K$ and $\eta_\pi$ decreases
with the introduction of $K-\pi$ interaction but the 
$\eta_K$ remains higher in magnitude than $\eta_\pi$ as the case
when the $K-\pi$ interaction is neglected.

In the context of phenomenological direction, the contribution of kaon component
at high temperature may only be relevant and our studies show that
%---------------------------------------------------------------------------
kaons play a crucial role in both electrical conductivity and
shear viscosity (and possible for other transport coefficients as well) at that high
temperature region.

%In phenomenological context, we may have to consider only the high $T$ kaon component. 
%Our results suggest that in the vicinity
%of the quark-hadron transition temperature, electrical conductivity of kaon
%component is almost equal with that of pion, while for shear viscosity case,
%kaons has larger contribution than pions.

%%%%%%%%%%%%%%%%%%%%%%%%%%%%%%%%%%%%%%%%%%%%%%%%%%%%%%%%%%%%%%%%%%%%%%%%%%%
{\bf Acknowledgment :} Sabyasachi Ghosh is supported from UGC Dr. D. S. Kothari Post Doctoral Fellowship under
grant No. F.4-2/2006 (BSR)/PH/15-16/0060.  Snigdha Ghosh (CNT project No. 3/2/3012/VECC/R\&D-I/1928)  
and MR are grateful to Department of Atomic Energy, Govt of India for financial support.
%
% 
%%%%%%%%%%%%%%%%%%%%%%%%%%%%%%%%%%%%%%%%%%%%%%%%%%%%%%%%%%%%%%%%%%%%%%%%%%%

\end{document}